%% file: ssd.tex
\documentclass[letterpaper,twocolumn,10pt]{article}
\usepackage{usenix,epsfig,endnotes}
\usepackage{url}
\usepackage{amsmath}
\usepackage{algpseudocode}	
\usepackage{algorithm}

\begin{document}

\date{}

\title{\Large \bf Serifos: Workload Consolidation and  Load Balancing for SSD Based Cloud Storage Systems}

\author{
{\rm Zhihao Yao}\\
Computer \& Information Technology\\
Purdue University
\and
{\rm Ioannis Papapanagiotou}\\
Cloud Database Engineering\\
Netflix
\and
{\rm Rean Griffith}\\
VMware
} 

\maketitle


\subsection*{Abstract}
Achieving high performance in virtualized data centers requires both deploying high throughput storage clusters, i.e. based on Solid State Disks (SSDs), as well as optimally consolidating the workloads across storage nodes. Nowadays,
 the only practical solution for cloud storage providers to offer guaranteed performance 
is to grossly over-provision the storage nodes. The current workload scheduling mechanisms used in production do not have the intelligence to optimally allocate block storage volumes based on the performance of SSDs. In this paper, we introduce Serifos, an autonomous performance modeling and load balancing system designed for SSD-based cloud storage. Serifos takes into account the characteristics of the SSD storage units and constructs hardware-dependent workload consolidation models. Thus Serifos is able to predict the latency caused by workload interference and the average latency of concurrent workloads. Furthermore, Serifos leverages an I/O load balancing algorithm to dynamically balance the volumes across the cluster.

Experimental results indicate that Serifos consolidation model is able to maintain the mean prediction error of around 10\% for heterogeneous hardware. 
As a result of Serifos load balancing, we found that the variance and the maximum average latency are reduced by 82\% and 52\%, respectively. The supported Service Level Objectives (SLOs) on the testbed improve 43\% on average latency, 32\% on the maximum read and 63\% on the maximum write latency.



\input{intro}

\input{relatedwork}

\input{background}

\input{modeling}

\input{system}

\input{evaluation}

\section{Conclusion and Future Work}

In this paper, we presented Serifos, an autonomous performance modeling and load balancing system designed for SSD based cloud storage infrastructures.

The first key contribution of Serifos is a workload consolidation model for predicting the average latency for multiple concurrent workloads running on a shared SSD device. After an exhaustive evaluation of workload parameters, our findings indicate that the write ratio and block size of a workload have a strong correlation to the
average latency. Moreover, the host-wide average latency can be predicted by the sum of workloads’ write ratios and the sum of workloads’ block sizes on a shared host. The six consolidation models built by Serifos are able to precisely predict the average latency of any combination of workloads and are used as guidelines for I/O load balancing.

Second, we integrate a load balancing engine with Serifos. The load balancer has a goal to optimize the system-wide latency performance by dynamically migrating workloads across storage servers. With the help of accurate consolidation models, the migrations recommended by the load balancer deliver a significant improvement in the latency performance of I/O workloads running in the backend storage systems.

We evaluated Serifos on an actual SSD deployment. Our experimental results indicate that Serifos consolidation model is able to maintain the mean prediction error around 10\% for heterogeneous hardware. The load balancing engine enhanced the average latency performance of the testbed managed by OpenStack Cinder by 82\%.
Furthermore, the supported performance SLOs in 99\% service time are reduced by 43\% on average latency, 32\% on the maximum read, and 63\% on the maximum write latency. 

In the future, we plan to consider the migration cost in the I/O load balancing algorithm. Some special cases such as migrating a large volume with a small amount of improvement needs to be carefully considered.

{\footnotesize \bibliographystyle{acm}
\bibliography{ssd}}


\end{document}

%% file: intro.tex
 \section{Introduction}

Cloud computing is a popular paradigm for the dynamic provisioning of computing and storage resources deployed on virtualized infrastructure. There is a constant need for higher performance (e.g. latency) by the virtualized storage infrastructure. As the cost of SSDs gradually decreases, cloud storage providers can provide higher throughput, lower access times, and lower power consumption than the legacy storage systems which are based on spinning disks. For example, Amazon Web Services \cite{aws} and Rackspace \cite{rack} have recently adopted SSDs as the main storage backend to support their elastic block storage service. In some cases, customized SSDs \cite{ouyang2014sdf} are used to meet the critical requirements of I/O performance. 

The problem of scheduling a storage volume in a multi-tenant storage cluster 
is more complicated than Virtual Machine (VM) scheduling, which focuses
primarily on slicing CPU and memory resources across VMs. First, in SSD-based cloud storage systems, the internal characteristics of SSDs, e.g., the activity of garbage collection or the allocation of over-provisioning space \cite{kim2015towards},
cannot be exposed to the virtualization layer. Second, the workload interference in a shared storage medium, especially SSDs, may decrease the performance further.
Finally, workload access patterns are not known in advance. Some existing workload schedulers such as the default available capacity scheduler in OpenStack block storage service, Cinder \cite{cinder}, only take into account the available space of the storage hosts. The lack of performance considerations in scheduling decisions can lead to performance degradation and Service Level Objective (SLO) violations. 

For compute resources (CPU and Memory), live migration of VMs is used to deal 
with unexpected workload variations and to alleviate hotspots~\cite{clark2005live, wood2007black}. However, storage volume migration is more complex and costly~\cite{mashtizadeh2011design} as the resources are rather shared and cannot be sliced. For example, workload interference can cause the cluster to underperform, an incorrect migration may affect other tenants, the time to perform a migration is dependent on external factors (e.g. network throughput), and the data may expire or change during migration. Hence in block storage migration the performance of the backend cluster must be predicted to ensure the benefit prior to executing a migration operation.

With the goal of equalizing the average (and higher percentile) latencies on every SSD based storage host, Serifos dynamically consolidates workloads so that no host is overloaded or underloaded in terms of I/O resources. Serifos extends other I/O load balancing systems like Romano \cite{park2012romano} and Basil \cite{gulati2010basil} to include capabilities such as predicting the aggregate latency for any combination of workloads, and dynamically allocating workloads to achieve a global balanced state of I/O. We use linear regression to construct the workload consolidation models based on the write ratio and the block size of the I/O workloads. In Serifos, the consolidation models are able to precisely predict the host-wide latency for heterogeneous hardware. An I/O load balancing algorithm is integrated into Serifos to achieve the global balanced state.

More specifically, Serifos makes the following contributions:
\begin{itemize}
	\item Serifos is \textit{accurate} enough to predict the average latency for consolidated workloads on SSD based storage hosts. We create workload consolidation modes according to the performance characteristics of SSD device to ensure the prediction accuracy. 
	\item Serifos is \textit{flexible} enough to manage different I/O access patterns on heterogeneous hardware by building machine dependent models. 
	\item Serifos proves to be \textit{robust} at balancing both at the average and the 99th percentile of the read and write latency across the storage infrastructure. Moreover, Serifos demonstrates the ability to support better performance SLOs. 
\end{itemize}

The paper is structured as follows: In Section 2, related work is briefly introduced. Section 3 presents the background on linear regression techniques and the scheduling in OpenStack Cinder, the performance baseline in our work. Section 4 shows the analysis of the performance characteristics of SSD devices and the workload consolidation model we developed. The design of Serifos' modeler and I/O load balancer is presented in Section 5 followed by evaluation results in Section 6. Finally we conclude our work in Section 7.

%% file: relatedwork.tex
\section{Related Work}

Workload management has been an active topic in academia and industry since the emergence of virtualization and cloud technologies. Prior work has focused on a variety of related problems such as performance modeling and prediction, workload consolidation, and SLO optimization.

Starting from the modeling of SSDs, Agrawal et al. \cite{agrawal2008design} studied the hardware layer design tradeoffs that impact the performance of SSDs through a trace driven simulator. The authors pointed out that SSD performance and lifetime is workload-sensitive. Chen et al.  \cite{chen2009understanding} and Hu et al. \cite{hu2009write} extended the aforementioned work and talked about how the SSD performance is affected by fragmentation, garbage collection, and write amplification. Subsequently, Huang et al. \cite{huang2011performance} characterized the performance patterns and built black-box models for one workload running on SSDs to predict performance. The authors indicated that the performance of a single workload on SSDs is predictable, but the mean relative error (MRE) on the extended model (8 parameters) could be as high as 20\%. On the contrary, by using two workload parameters, Serifos' performance model for one workload decreases the MRE to 5\% and 7\% on two SSD devices, respectively. Serifos also extends the performance modeling to a shared environment with many concurrent workloads. Finally, we test Serifos with actual enterprise level SSDs.



Some other studies have looked into developing high level scheduling algorithms and workload management systems. Romano \cite{park2012romano} is a storage load balancing system designed to optimize the latency performance of traditional spinning disk based virtualized datacenters. It constructs a performance model via linear regression to predict the average latency of workloads, and subsequently performs I/O load balancing. Romano builds on top of systems like Pesto \cite{gulati2011pesto} and Basil \cite{gulati2010basil}. Due to the fundamental differences in the internal mechanisms of spinning disks and SSDs, such as the seek time or garbage collection, Romano, Pesto and Basil are not suitable for predicting the performance of workloads on SSDs. Furthermore, one consolidation model for two workloads is built in these legacy systems. This may result in large prediction error when scaling to many concurrent workloads.  
The goal of Serifos is similar to these systems and extends them to SSDs. Hence the performance pattern is different from spinning disks, which causes the workload parameters used in the modeling steps and the model to be different. 
Six different consolidation models created in time intervals similar to Romano provide accurate prediction at scale.

Finally, there are management frameworks focusing on providing better SLOs to block storage consumers. Some, e.g.,~\cite{yao2014sla, yaomulti}, proposed near optimal scheduling algorithms that considered multi-dimensional resources to enable performance-SLO management capabilities for cloud storage backend systems. PriorityMeister \cite{zhu2014pm} focused on providing end-to-end tail latency QoS to meet performance SLOs. By creating an enforcer daemon on the storage and network side, SLOs could be respected. In a similar fashion, Cake \cite{wang2012cake} enforced SLOs by limiting the outstanding I/Os (OIOs) at multiple tiers of the storage system. Unlike Serifos, none of these frameworks considered the characteristics of the underlying hardware, which limited their effects in heterogeneous environments.

In short, Serifos is the first I/O load balancing system, which is specifically designed for SSD-based shared storage infrastructure. Based on the observation of SSD performance characteristics, six consolidation models per hardware type are able to provide precise latency predictions. The I/O load balancing algorithm in Serifos can deliver significant improvements in system-wide performance while supporting better SLOs. We present our observation and design detail in the next two sections.

%% file: background.tex
\section{Background}

\subsection{Linear Regression}

In statistics, regression is an approach for modeling the relationship between a dependent variable $P$ and one or more independent variables denoted $X$. A generic model that predicts $P$ given $X$ can be expressed as: 
\begin{equation}
P = f(X, \beta) + \varepsilon
\end{equation}
In this work, we mainly use the linear regression shown in Equation \ref{lm} to construct the models because of its lower computation complexity and reasonable model accuracy. 

\begin{equation}
\label{lm}
P  = \beta_0 + \beta_1x_1 + \beta_2x_2 + ... + \beta_nx_n + \varepsilon
\end{equation}
In order to evaluate the goodness-of-fit for linear regression model, we employ the R-squared metric ($R^{2}$) for simple linear models and adjusted R-squared for multiple linear regression. R-squared is a statistical measure of how close the data are to the fitted regression line and it indicates the percentage of the response variable variation that is explained by a linear model. R-squared values fall between 0 and 100\%. In general, the higher the R-squared, the better the model fits the data. The adjusted R-squared is a modified version of R-squared that accounts for the number of predictors in the model. The adjusted R-squared value increases only if the new term improves the model more than would be expected by chance.

\subsection{OpenStack Cinder}
\label{subsec:oscinder}

OpenStack \cite{openstack} is the most heavily used open source cloud computing platform for public and private clouds. It consists of multiple service components providing different Infrastructure as a Service (IaaS) abstractions. The block storage service in OpenStack \cite{cinder}, code name Cinder, provides a service for managing virtualized block storage devices (i.e. volumes) on various storage backend systems. The cinder-scheduler daemon is responsible for scheduling virtual volumes to physical storage hosts. 

\begin{figure}[t]
	\centering
		\includegraphics[width=0.5\textwidth]{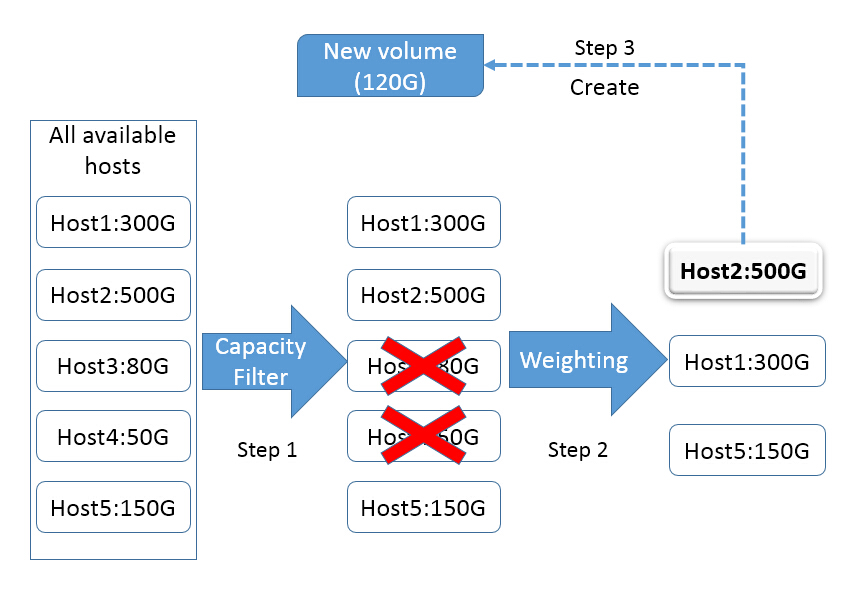}
	\caption{Volume scheduling in OpenStack cinder}
	\label{fig:cindersche}
\end{figure}

The default scheduling algorithm in Cinder is called \textit{Available Capacity} \cite{cindersche}. It is based on a filtering/weighting procedure as shown in Figure \ref{fig:cindersche}. The \textit{Available Capacity} scheduler works in three steps. When a new volume request comes in, the scheduler first filters out the hosts which have insufficient resources (e.g. capacity) to meet the needs of the request. Then the scheduler calculates a weight for each of the remaining hosts based on the available capacity, and sorts them based on this weight in a increasing order. The host which has the least weight is chosen as the best candidate to serve the new request. In this work, we use the default \textit{Available Capacity} scheduler of Cinder as the performance baseline, since it is a standard service in OpenStack and many clients will keep the default configuration on their deployment.

%% file: modeling.tex
\section{SSD Performance Modeling}

Exploring the workload's parameters and corresponding latency performance on SSDs provides the essential support for the optimized consolidation algorithm. Previous work by Agrawal et al. \cite{agrawal2008design} and Chen et al. \cite{chen2009understanding} has pointed out that the performance of SSDs is highly workload dependent. Similar to our work, prior studies have focused on the performance model of SSD devices. In our work, we validate some of these findings as well as extend them to a shared environment based on next generation enterprise-level SSDs. Hence, in this section we present our observations on the SSD performance characteristics and consolidation model for concurrent workloads.

\subsection{Methodology}

\begin{table}[t]
	\centering
		\begin{tabular}{|c|c|}
			\hline
				Option & Note \\
			\hline
				direct=1 & Bypass OS cache \\
			\hline
				ioengine=libaio & Linux native asynchronous io library \\
			\hline
				refill\_buffers=1 & Refill the I/O buffers on every submit \\
			\hline
				runtime=60 & Every job lasts 1 minute \\
			\hline
				iodepth=8 & I/O concurrency for one workload \\
			\hline	
		\end{tabular}
	\caption{Fio configuration values}
	\label{tab:Fioopt}
\end{table}

We use three variables as recommended by the Storage Networking Industry Association (SNIA) \cite{sniassd} to represent the access patterns of the workloads on SSDs: 
\begin{itemize}
\item Write ratio ($W$): the percentage of write requests in total requests.
\item Randomness ($X$): the percentage of random access I/Os in total requests.
\item Block size ($S$): the number of bytes transferred to/from the storage device. The default unit is kilobyte (KB).
\end{itemize}

We employed a synthetic workload generator, Fio \cite{fio}, to generate specific I/O access patterns based on these parameters ($W$, $X$ and $S$). We use the Linux native asynchronous I/O library as the default I/O generation engine. To remove the cache effects, the OS cache is bypassed by enabling direct I/O. Some other Fio global configuration values are shown in Table \ref{tab:Fioopt}. These configurations are referenced from several prior studies~\cite{huang2011performance, lu2015vfair, tkperf}. We measure the average latency and use it as the main performance metric -- units are microseconds ($\mu$s) unless otherwise indicated.

\begin{table}[t]
	\centering
		\begin{tabular} {|c|c|c|}
			\hline
				Server name & SSD1 & SSD2 \\
			\hline
				Model & \multicolumn{2}{|c|}{Dell R430} \\
			\hline
				CPU & \multicolumn{2}{|c|}{Intel Xeon E5-2620 v3 2.4GHz} \\
			\hline
				Memory & \multicolumn{2}{|c|}{8GB} \\
			\hline
				SSD model & DC S3500 & DC S3610 \\
			\hline
				SSD capacity & 480GB & 400GB \\
			\hline
				OS & \multicolumn{2}{|c|}{Ubuntu Server 14.04 LTS} \\
			\hline
		\end{tabular}
	\caption{Testbed server configuration.}
	\label{tab:serverspec}
\end{table}

Our testbed consists of two models of SSDs, the Intel DC S3500 series released in 2013, and the Intel DC S3610 series released in 2015.
Table \ref{tab:serverspec} describes the detailed configurations of the servers.  Each of the servers was equipped with a set of these drives, but we only use one drive to perform the I/O tests. We used the other drive as the OS drive so that the Fio-issued requests are only sent to the test drive. Instead of using the raw device (e.g. /dev/sdb), we use LVM (Logical Volume Manager) to create block volumes and connect each volume to the Fio process. We choose LVM to mimic
real-world deployments since it is the default volume driver in OpenStack Cinder.

An important first step in SSD performance measurements is {\em aging} the device \cite{kim2015towards}. Typically a fresh SSD exhibits a transient period of elevated performance, which evolves to a stable performance state relative to the workload being applied. It is critical to ensure the SSD under test workload is running at the steady state to collect meaningful results. A method \cite{sniassdtest}, suggested by SNIA, is applied to age the device before running any experiments so that the SSD state at the start of every experiment is in a steady and consistent state. First, the device is purged by issuing the ATA secure erase command. Then we age the device by running a special workload issuing 128KB sequential write I/O requests to the entire logical block addresses for 2X (twice) the user capacity. Each experiment is performed five rounds and the result of the fifth round is used as valid result.

\subsection{Device Model}

We build a device model to describe the correlation between the significant workload parameters and the average latency of a workload running on the SSDs. We employ linear regression and analysis of variance (ANOVA) to determine whether the three workload parameters are effective in predicting the average latency. 

The test vector of each parameter used to describe all possible access patterns of an I/O workload are defined as:
\begin{equation} \label{writev}
	W = \left\{ 0\%, 10\%, 20\%, 30\%, \dots 90\%, 100\% \right\} 
\end{equation}
\begin{equation} \label{ran}
	X = \left\{ 10\%, 20\%, 30\%, \dots 90\%, 100\% \right\} 
\end{equation}
\begin{equation} \label{bsv}
	S = \left\{ 4, 8, 16, 32, 128, 256 \right\}
\end{equation}
This test vector consists of 660 different access patterns in total for one workload. Note that we exclude the $X=0\%$ which represents a $100\%$ sequential workload. This is due to the fact that pure sequential workloads have completely different performance patterns on an SSD compared to random workloads in terms of latency. Similar observations were also made by \cite{agrawal2008design, chen2009understanding, huang2011performance}. In fact, even for a spinning disk based storage systems, an 100\% sequential workload would not provide any benefit to performance modeling other than to add a large relative error \cite{gulati2010basil}.


\begin{table}[t]
	\centering
		\begin{tabular}[0.5\textwidth]{ |c||c|c||c|c| }
			\hline
				 & \multicolumn{2} {|c||} { SSD1 } & \multicolumn{2} {|c|} { SSD2 } \\
			\hline
				Adj. $R^2$ & \multicolumn{2} {|c||}{ 0.98 } & \multicolumn{2}{|c|}{ 0.991 } \\
			\hline
				factor & Coef & p-value & Coef & p-value \\ 
			\hline
				INTCP & -852.76 & 2.42e-15 & -346.32 & 2.6e-5\\
			\hline
				$W$ &  32.596 & 2.68e-36 & 27.815 & 1.19e-42 \\
			\hline
				$X$ & 3.367 &  1.13e-5 & \textbf{1.8305} & \textbf{0.439} \\
			\hline
				$S$ & 29.382 & 3.88e-125 & 21.204 & 3.01e-282 \\
			\hline
				$W:X$ & -0.1645 & 7.68e-16 & -0.0424 & 0.00676 \\
			\hline
				$W:S$ & 0.2028 & 1.42e-138 & 0.0832 & 6.62e-146 \\
			\hline
				$X:S$ & -0.051 & 2.31e-13 & \textbf{0.0055} & \textbf{0.0529} \\
			\hline
				$W^2$ & -0.246 & 3.3e-30 & 0.2761 & 3.05e-56 \\
			\hline
				$X^2$ & \textbf{0.045} & \textbf{0.069} & \textbf{0.0001} & \textbf{0.994} \\
			\hline
				$S^2$ & -0.0108 & 0.0095 & -0.0041 & 1.16e-11  \\
			\hline
		\end{tabular}
	\caption{The polynomial model for two types of storage server. Terms with high p-values are bolded. INTCP and Coef are the abbreviation for intercept and coefficient respectively.}
	\label{tab:quaddevicemodel}
\end{table}

We first apply polynomial regression on the measured latency to obtain the most accurate model. The polynomial model contains an intercept, linear terms, interactions and squared terms. Romano \cite{park2012romano} found that sometimes the interaction and squared terms show enough impact in the model for spinning disks based storage infrastructure. The polynomial regression results of two SSD servers are shown in Table \ref{tab:quaddevicemodel}. Both coefficients and p-values are captured. We derive the p-value from the ANOVA test at the 95\% confidence level. It tests the null hypothesis that the coefficient of a term is equal to zero (no effect). If the p-value of a term is greater than 0.05, it suggests that the null hypothesis is accepted and the changes in the predictor are not associated with the changes in the response, and therefore the predictor can be removed from model safely. Terms with high p-value ($>$ 0.05) are shown in \textbf{bold} font in the table.

In both models, we notice that the interaction terms and squared terms have very small coefficient. Yet some of them are significant (p-value $<$ 0.05). This indicates that these terms have minimal impact on the average latency on SSD device. Based on this observation, we decide to apply linear regression to build a device model without interaction and squared terms.

\begin{table}[t]
	\centering
		\begin{tabular}[0.5\textwidth]{ |c||c|c||c|c| }
			\hline
				 & \multicolumn{2} {|c||} { SSD1 } & \multicolumn{2} {|c|} { SSD2 } \\
			\hline
				Adj. $R^2$ & \multicolumn{2} {|c||} { 0.94 } & \multicolumn{2} {|c|} { 0.975 } \\
			\hline
				factor & Coef & p-value & Coef & p-value \\ 
			\hline
				INTCP & -421.19 & 2.13e-6 & -351.13 & 1.722e-7\\
			\hline
				$W$ &  13.959 & 3.1e-39 & 9.228 & 1.584e-31 \\
			\hline
				$X$ & -2.934 &  0.0009 & \textbf{0.4722} & \textbf{0.57} \\
			\hline
				$S$ & 33.968 & 0 & 23.644 & 0 \\
			\hline
		\end{tabular}
	\caption{The linear model for two types of storage server. 
      Terms with high p-values are bolded.}
	\label{tab:lineardevicemodel}
\end{table}


Table \ref{tab:lineardevicemodel} describes the device model in linear form. The adjusted R-squared values of both models are 0.94 and 0.975 respectively. Therefore, the accuracy of both models does not decrease much due to the exclusion of interaction and squared terms. The linear terms are able to precisely describe the device model. Another observation from the linear models is that randomness is not a significant factor to the average latency of the $SSD2$ server model and has a very small coefficient in the $SSD1$ server model. 
To illustrate this fact, we analyze the effect on average latency of each workload parameter in detail. When randomness changes from 10\% to 100\%, the average latency only varies about 0.3ms, given that all other parameters are held constant, whereas the effect of block size is nearly 9ms. Write ratio can affect the average latency by almost 2ms. Hence the randomness may not be a important factor in terms of average latency of SSD device. The linear models without randomness are shown at Equation \ref{ssd1_2p} and Equation \ref{ssd2_2p}. 
%

\begin{eqnarray}
\label{ssd1_2p}
\begin{split}
L_{avg\_SSD1} = -621.06 + & 12.96W + 33.97S \\
          & (Adj.R^2 = 0.94)
\end{split}
\end{eqnarray}
\begin{eqnarray}
\label{ssd2_2p}
\begin{split}
L_{avg\_SSD2} = -325.59 + & 9.23W + 24.64S \\
          & (Adj.R^2 = 0.975)
\end{split}
\end{eqnarray}

The adjusted R-squared values do not change in the simplified models that exclude randomness. Therefore we can conclude that only write ratio and block size have significant contributions to the average latency of the non-sequential workload running on SSD.  

\subsection{Workload Consolidation Model}

In a cloud storage infrastructure, the resources of a single physical host are shared among several concurrent workloads. In order to predict the host-wide average latency, we need to investigate the performance patterns of consolidated workloads.

\begin{table}[t]
	\centering
		\begin{tabular}[0.5\textwidth]{ |c||c|c||c|c| }
			\hline
				 & \multicolumn{2} {|c||} { SSD1 } & \multicolumn{2} {|c|} { SSD2 } \\
			\hline
				Adj. $R^2$ & \multicolumn{2} {|c||} { 0.972 } & \multicolumn{2} {|c|} { 0.986 } \\
			\hline
				factor & Coef & p-value & Coef & p-value \\ 
			\hline
				INTCP & 583.36 & 2.285e-14 & -275.38 & 8.929e-11\\
			\hline
				$W_1$ &  -4.105 & 4.667e-8 & 4.426 & 8.3927e-20 \\
			\hline
				$W_2$ & -3.903 &  3.77e-7 & 4.495 & 2.326e-20 \\
			\hline
				$S_1$ & 23.26 & 0 & 19.663 & 0 \\
			\hline
				$S_2$ & 23.164 & 0 & 19.713 & 0 \\
			\hline
		\end{tabular}
	\caption{The consolidation models for two workloads }
	\label{tab:con2model}
\end{table}

The methodology is similar to the device model steps. However only the write ratio ($W$) and block size ($S$) are used to describe a workload. First, we examine the model for two workloads in which the write ratio and block size are the same as in Equation \ref{writev} and \ref{bsv}. The total number of access patterns in the test set is 4356 or ${(11 \times 6)}^2$. The linear models of two servers are shown in Table \ref{tab:con2model}. Both models fit the data very well according to their adjusted R-squared values. This proves that the write ratio and block size can effectively be used to predict average latency in the shared environment.

From the consolidation models, we notice that the coefficients of write ratio, $W_1$ and $W_2$, are very close, so are the coefficients of block size, $S_1$ and $S_2$. So there is an opportunity to combine the corresponding parameters of two workloads in the model. We calculate the sum of the write ratio and the sum of the block size, and apply them to build the consolidation model in another form, shown as follows:
\begin{eqnarray} \label{ssd1_con2}
\begin{split}
L_{avg\_SSD1} = & -583.36 + -4.02(W_1 + W_2) \\
          & + 23.21(S_1+S_2) \quad (Adj.R^2 = 0.975)
\end{split}
\end{eqnarray}
\begin{eqnarray} \label{ssd2_con2}
\begin{split}
L_{avg\_SSD2} = & -275.38 + 4.46(W_1 + W_2) \\
          & + 19.68(S_1+S_2) \quad (Adj.R^2 = 0.986)
\end{split}
\end{eqnarray}

These models do support the conclusion that there is a strong correlation between the average latency and the workload parameters in a sum format. To verify this observation, we scale the number of concurrent workloads to 3 and 4. We reduce the running time of one test from 1 minute to 30 seconds so that the whole set of tests can be done in couple of days. The linear models of 3 workloads are: 
\begin{eqnarray} \label{ssd1_con3}
\begin{split}
& L_{avg\_SSD1} =  419.21 + -2.51(W_1 + W_2 + W_3) \\
          & + 24.987(S_1+S_2 + S_3) \quad (Adj.R^2 = 0.953)
\end{split}
\end{eqnarray}
\begin{eqnarray} \label{ssd2_con3}
\begin{split}
& L_{avg\_SSD2} =  34.233 + -0.357(W_1 + W_2 + W_3) \\
          & + 21.561(S_1+S_2 + S_3) \quad (Adj.R^2 = 0.972)
\end{split}
\end{eqnarray}
and 4 workloads:
\begin{eqnarray} \label{ssd1_con4}
\begin{split}
& L_{avg\_SSD1} =  179.65 + \\
          & -1.621(W_1 + W_2 + W_3 + W_4) \\
          & + 24.684(S_1+S_2 + S_3+ S_4) \quad (Adj.R^2 = 0.941)
\end{split}
\end{eqnarray}
\begin{eqnarray} \label{ssd2_con4}
\begin{split}
& L_{avg\_SSD2} =  -302.78 + \\
          & -1.121(W_1 + W_2 + W_3 + W_4) \\
          & + 21.937(S_1+S_2 + S_3+ S_4) \quad (Adj.R^2 = 0.983)
\end{split}
\end{eqnarray}
The coefficients of all workloads are still close enough to be combined into a sum term. The adjusted R-squared values of all models indicate a good fit on measured data. Therefore, the workload consolidation model can be expressed as:
\begin{equation} \label{lcon}
	L = \beta_0 + \beta_1\sum\limits_{i=1}^n W_{i} + \beta_2\sum\limits_{i=1}^n S_{i} + \varepsilon
\end{equation}
where $n$ is the number of concurrent workloads. According to this model, we are able to apply it to predict the host-wide average latency based on the write ratio and block size of every concurrent workload on the shared physical storage host. Moreover, the coefficient of the sum of workloads' block sizes is much larger than the coefficient of the sum of write ratios in all above models. Hence we can conclude that the block size is the major latency predictor in the consolidation model compared with write ratio, although both of them show significance in the model.

%% file: system.tex
\section{System Design of Serifos}

Based on the workload consolidation model, we design Serifos, an I/O load balancing system for SSD based shared storage infrastructure. The goal of Serifos is to optimize the system-wide latency performance. Serifos is designed to be integrated into OpenStack as a dynamic scheduling plug-in. There are two major components in the Serifos: the \textit{Modeler} and the \textit{Load-balancer}. 

\textbf{Modeler: }Before a new server is deployed in a production environment, the modeler runs five synthetic test sets to generate six workload consolidation models to predict the average latency, including five basic models and one extended model. Each test set consists of a combination of a fixed number (1-5) of workloads with different access patterns. One basic workload consolidation model is generated based on collected average latency and related workload combinations from one test set. Five basic models derived from five test sets are responsible for providing the prediction for 1-5 workloads respectively. The extended model is built based on the aggregated measurements of all five test sets, and is applied when there are more than five concurrent workloads. The intercept and coefficients of models will be stored in a JSON file which will be used by the \textit{Load-balancer}. In order to reduce the time to build the model, we used a simplified version of the test vectors, shown in Equation \ref{m_writev} and \ref{m_bsv}. The evaluation results in Section 6 show that such simplification do not affect the prediction accuracy of all consolidation models.
\begin{equation} \label{m_writev}
	W = \left\{  25\%, 50\%, 75\% \right\} 
\end{equation}
\begin{equation} \label{m_bsv}
	S = \left\{ 4, 8, 32, 128\right\}
\end{equation}

It is possible that the write ratio is not significant (p-value $>$ 0.05) in the standard form of consolidation model with simplified test vectors. In such cases, the non-significant factor is removed from the model and corresponding coefficient is set to 0. Then the consolidation model is re-built based on the significant factors.

We carefully evaluated the test cases in the test set. We found that some tests of multiple workloads are actually repeated because the order of workloads do not play a role in terms of performance. For example, the test pattern [w1(25\%, 8k), w2(50\%, 128k)] and [w1(50\%, 128k), w2(25\%, 8k)] are the same cases. Such replicated test cases are removed from the test set by generating the mathematical combinations (with replacement) of the test vectors of workloads. Comparing with the calculation of the Cartesian product of the test vectors, the time needed for running all test sets reduces from several weeks to only one and half days (including device preconditioning).

Some previous systems, such as Romano \cite{park2012romano} and Basil \cite{gulati2010basil}, only create one consolidation model of two aggregated workloads. This model is recursively applied for predicting performance when there is more workloads. In Serifos, with the benefit from test vector simplification and test set de-duplication, we are able to run more complex tests for more workloads within similar time consumption. Five basic models allow very accurate predictions on the basis of real measurements. The idea of the extended model is based on the observations that the block size is the most determined parameter to the average latency and the coefficients of the block size term in Equation \ref{ssd1_con2}-\ref{ssd2_con4} are very close ($\sim 24 for SSD1, \sim 21 for SSD2$). We make a reasonable assumption that this trend can scale to more concurrent workloads. Therefore, we build an extended model for more than five workloads depending on the aggregated measurements. The evaluation results in Section 6.1 verify our assumption and prove that the extended model is able to precisely predict the average latency.

\textbf{I/O load Balancer: } The I/O load balancer is the major component in the system. It aims to equalize the average latency across multiple storage servers and to reach the global balanced state by re-allocating the existing workloads. Based on the consolidation models built by the \textit{Modeler} module, the average latency can be predicted before performing any workload migration to ensure performance benefits.    

Our core balancing algorithm is inspired by the Best Fit Decreasing (BFD) approximation of the well-known NP-hard bin packing problem. In the classical problem, $N$ items with different sizes are placed in a set of bins. The goal is to pack the items in as few bins as possible. Among a various of bin packing algorithms, BFD is able to achieve the best asymptotic worst-case performance \cite{johnson1974worst}. The BFD algorithm is an offline approximate approach with the time complexity of $O(n\log n)$ that sorts items at the beginning according to their size in a decreasing order so that the large items can be processed first. Then each item is placed in the best, i.e. tightest, bin to minimize the empty space. In our scenario, the corresponding attribute of item size is the block size of a workload. Comparing with write ratio, the block size has a much greater impact and is the decisive factor for average latency based on the observation on device model and consolidation model. The best is also defined as the lowest average latency among all predictions.      
   
\begin{algorithm}[t]
	\caption{I/O Load Balancing Algorithm}
	\label{algo1}	
	\renewcommand{\algorithmicforall}{\textbf{foreach}}
	\begin{algorithmic}
		\Procedure{load\_balance}{$wkld\_list, host\_list$}
			\State $wkld\_list$.sortDecreasing($wkld.blockSize$)
			\ForAll {$wkld$ in $wkld\_list$}
				\State $best\_host$ = $None$
				\State $best\_perf$ = $MAX\_PERF$
				\ForAll {$host$ in $host\_list$} 
					\If {$host.freeCapacity < wkld.size$}
						\State \textbf{continue}
					\EndIf
					\State $perf$ = PredictPerf($host$, $wkld$)
					\If {$perf < best\_perf$}
						\State $best\_host$ = $host$
						\State $best\_perf$ = $perf$
					\EndIf
				\EndFor
				\If {$best\_host \neq currentHost$}
					\State scheduleMigration($currentHost$, $best\_host$)
				\EndIf
			\EndFor
		\EndProcedure
		\Procedure{PredictPerf}{$host$, $wkld$}
			\State $model$ = loadModel($host.numOfWklds + 1$)
			\State $new\_wr\_sum$ = $host.wr\_sum$ + $wkld.wr$
			\State $new\_wr\_sum$ = $host.bs\_sum$ + $wkld.bs$
			\State $perf$ = $model$.predict($new\_wr\_sum$, $new\_bs\_sum$)
			\State \textbf{return} $perf$
		\EndProcedure
	\end{algorithmic}
\end{algorithm}

Algorithm \ref{algo1} outlines two functions running in the I/O load balancer. The main function $load\_balance$ takes two arguments as input: the workloads running in the system, and the available storage hosts. First, all workloads are sorted in decreasing order based on their block size, because the block size is the decisive factor to the average latency. Then the algorithm starts at the workload which has the largest block size. 
For a given workload, the function evaluates the predicted average latency of all available hosts if that workload is allocated on each host, under the assumption that the candidate hosts has enough free capacity to allocate that workload. Then the host having the lowest predicted average latency will be chosen as the best candidate to allocate the workload. If the current host of a workload is not the best host, the workload will be scheduled for a migration operation. The function $PredictPerf$ predicts the average latency after a new workload is placed on a host. The write ratio and block size of new workload are added to the corresponding sum of parameter of existing workloads on the host. After that the consolidation model is able to calculate the predicted average latency. 

It is obvious that the key factor impacting the balancing effect is the accuracy of the consolidation model. If the prediction error is not small enough, the balancer may make an incorrect choice of the best candidate host of a workload. As a consequence, the global balanced state can not be achieved, and I/O 
resources will be wasted on unnecessary workload migrations. This is also the reason we choose to build individual consolidation models for each number of workloads.

%% file: evaluation.tex
\section{Evaluation}

We implement a prototype of Serifos in Python to build the consolidation model and balance workload performance. We chose Python so that our system can be eventually committed to the OpenStack Cinder project and work with the OpenStack telemetry service (Ceilometer) to retrieve the necessary workload statistics. We also choose to develop a simple volume management framework based on LVM to support the modeler and I/O load balancer component for repeatable and quick evaluation. The default available capacity scheduling algorithm of Cinder, introduced in \ref{subsec:oscinder} was also implemented in the framework for comparison purposes. In this section, the prediction accuracy of the consolidation models and the I/O load balancing effects are evaluated through several experiments on a real testbed. There are five storage servers in the testbed, 3 of $SSD1$ type and 2 of $SSD2$ type (Table \ref{tab:serverspec}).

\begin{figure}[t]
	\centering
		\includegraphics[width=0.50\textwidth]{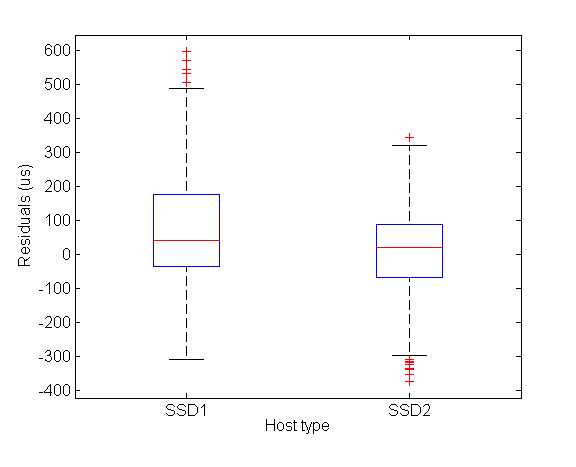}
	\caption{Residuals distribution for 500 tests. }
	\label{fig:Modelacc}
\end{figure}

\begin{figure*}[t]
	\centering
		\includegraphics[width=\textwidth]{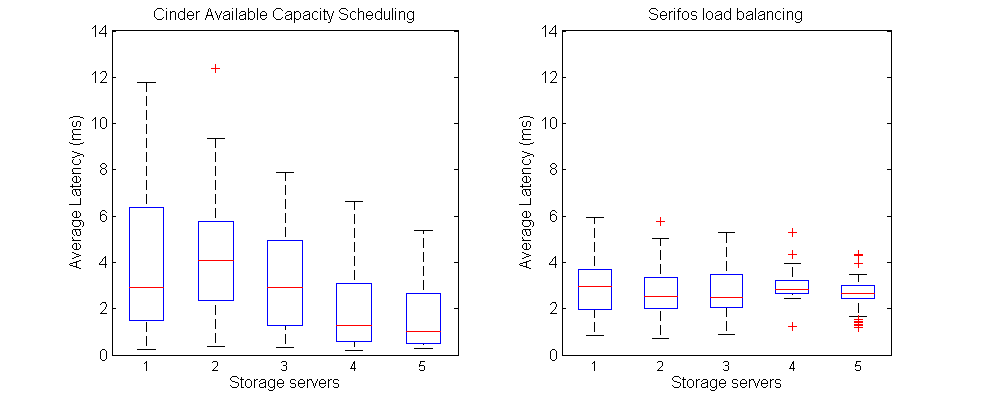}
	\caption{Average latency distribution on 50 test cases.}
	\label{fig:avgl_50}
\end{figure*}

A revised version of test vectors (Equations \ref{testwrite}-\ref{testsize}) is applied to generate representative workloads~\cite{ioworkload}. The default randomness for all access patterns is 100\%. Size $C$ is added as a new parameter of workload so that the LVM manager can create the storage volume of workload accordingly. The access pattern of a workload is randomly selected from these test vectors. In order to speed up the testing, workload migration is emulated by moving the Fio process from one server to another instead of actually copying data between different hosts. The attached volume is also deleted on the original server and re-created on the destination server. This helps us reduce the experimental time from weeks to days.
\begin{equation} \label{testwrite}
	W = \left\{ 5\%, 30\%, 50\%, 70\%, 95\%\right\} 
\end{equation}
\begin{equation} \label{testbs}
	S = \left\{ 4, 8, 16, 32, 64, 128, 256 \right\}
\end{equation}
\begin{equation} \label{testsize}
	C = \left\{ 30, 60, 90, 120 \right\} 
\end{equation}

\begin{table*}[t]
	\centering
		\begin{tabular}[\textwidth]{ |c||c|c|c|c||c|c|c|c| }
			\hline
				 & \multicolumn{4} {|c||} { SSD1 } & \multicolumn{4} {|c|} { SSD2 } \\
			\hline
				Workloads & INTCP & $W_{sum\_co}$ & $S_{sum\_co}$ & Adj. $R^2$ & INTCP & $W_{sum\_co}$ & $S_{sum\_co}$ & Adj. $R^2$ \\
			\hline
				1 & 113.44 & \textit{0} & 22.135 & 0.994 & 216.51 & -1.19 & 19.628 & 0.999 \\
			\hline
				2 & \textit{0} & \textit{0} & 24.497 & 0.988 & 42.669 & \textit{0} & 20.691 & 0.996 \\
			\hline
				3 & \textit{0} & \textit{0} & 24.714 & 0.981 & -86.634 & 0.533 & 21.339 & 0.995 \\
			\hline
				4 & 81.969 & \textit{0} & 23.587 & 0.977 & -188.26 & 0.907 & 21.729 & 0.994 \\
			\hline
				5 & \textit{0} & 0.578 & 23.919 & 0.98 & -133.83 & 0.519 & 21.906 & 0.994 \\
			\hline
				5+ & \textit{0} & 0.646 & 23.913 & 0.981 & -137.81 & 0.597 & 21.821 & 0.995 \\
			\hline
		\end{tabular}
	\caption{The consolidation models for two servers. If the coefficient is 0, it means the corresponding term is not significant (p-value $>$ 0.05) in the standard linear consolidation model. The model is re-built only based on significant terms and used in load balancing. }
	\label{tab:2servermodel}
\end{table*}

\subsection{Prediction of Consolidation Models}

\begin{table}[t]
	\centering
		\begin{tabular}{ |c||c||c| }
			\hline
				Model &  SSD1 &  SSD2 \\
			\hline
				1 & 7\% &  5\%  \\
			\hline
				2 & 12\% & 7\%  \\
			\hline
				3 & 12\% &  7\%  \\
			\hline
				4 & 10\% &  5\%  \\
			\hline
				5 & 9\% &  10\%  \\
			\hline
				5+ & 16\% & 8\% \\
			\hline
		\end{tabular}
	\caption{Mean relative error of the consolidation models}
	\label{tab:modelerror}
\end{table}
Before evaluating the prediction accuracy, the modeler is executed on one server of each type to construct the consolidation models. The regression results of five basic models (model 1-5) and one extended model (model 5+) for both types of server are shown in Table \ref{tab:2servermodel}. These models indicate that the block size on $SSD1$ server is the direct predictor for the average latency. The write ratio only plays a limited role on the consolidation model for 5 and more workloads. On $SSD2$ server, most models follow the general model form except the model for 2 workloads. However, the write ratio still has minimal impact compared with the block size. These models show that block size is the decisive predictor of the average latency again. Moreover, all of the models' adjusted R-squared values are very high which means the models fit the collected data very closely.

To examine the prediction accuracy of these models, we run 500 tests to evaluate five basic consolidation models (100 each) and 200 tests for extended model on both types of server. Each test on basic models contains a corresponding number of workloads with random access patterns. For the extended model, each test has random number of 6-10 workloads. The mean relative error (MRE) values of all models are listed in Table \ref{tab:modelerror}. Six models out of ten basic models have less than 10\% MRE and the worst MRE is only 12\%. The extended model of $SSD1$ has 16\% MRE while $SSD2$ server achieves an outstanding accuracy on the newest SSD device with only 8\% MRE. To deeply investigate the prediction accuracy, the distribution of basic models' residuals (difference between real and prediction) are presented in Figure \ref{fig:Modelacc}. The residuals of all basic models are calculated as the prediction error $\epsilon$. The median value of two servers are near 0 and the lower 50\% residuals are in the range of [-50, 200] and [-50, 100] approximately. Note that the unit of residuals is microsecond. 
These results demonstrate that the accuracy of Serifos consolidation models is robust enough to provide precise performance predictions to the I/O load balancing algorithm. Moreover, it validate the fact that our method of building the consolidation models is appropriate for SSD based storage.

\begin{figure*}[t]
	\centering
		\includegraphics[width=\textwidth]{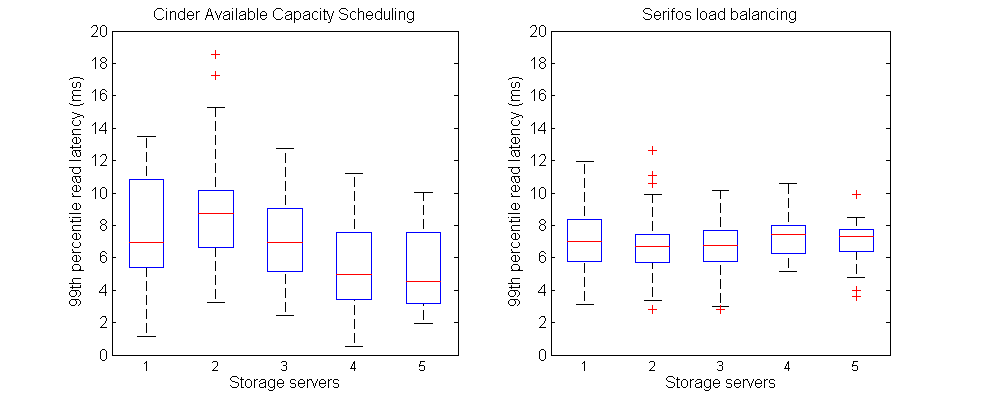}
	\caption{99th read latency distribution on 50 test cases.}
	\label{fig:read99}
\end{figure*}

\begin{figure*}[!htbp]
	\centering
		\includegraphics[width=\textwidth]{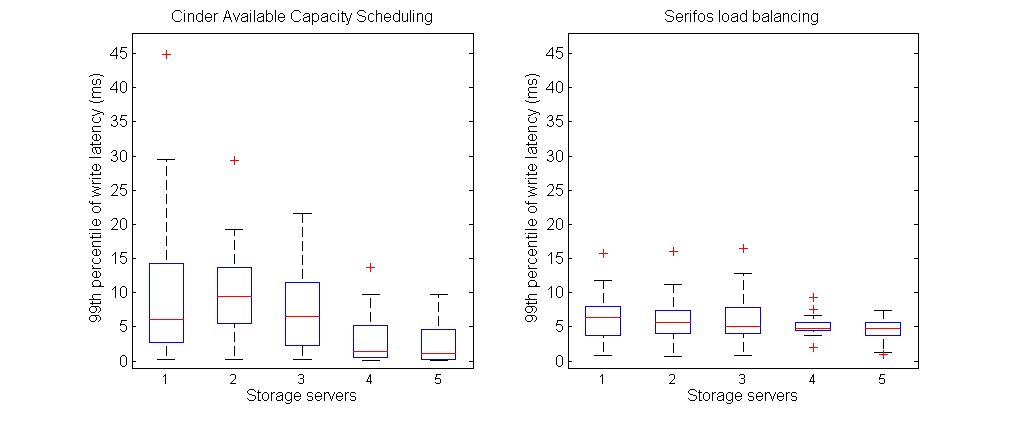}
	\caption{99th write latency distribution on 50 test cases.}
	\label{fig:write99}
\end{figure*}

\subsection{I/O Load Balancing}

\begin{figure}[!htbp]
	\centering
		\includegraphics[width=0.50\textwidth]{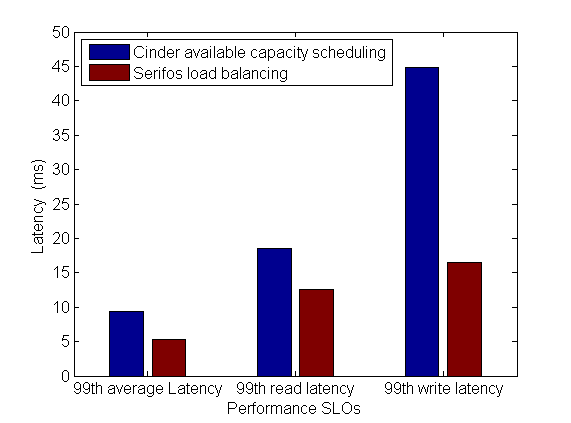}
	\caption{Performance SLO supported by OpenStack Cinder and Serifos.}
	\label{fig:slo}
\end{figure}

In this section, we randomly generate 10-16 workloads on the testbed which is managed by the available capacity scheduling algorithm in OpenStack Cinder. Then the load balancer is enabled to optimize the global latency on our testbed. The experiment is repeated 50 times.

Figure \ref{fig:avgl_50} shows the aggregated latency of Cinder scheduling and load balancing. In the results for the baseline Cinder scheduler, shown in the left graph, the first three hosts (type $SSD1$) are overloaded. Because the SSD capacity on $SSD1$ servers are larger than the $SSD2$, the Cinder scheduler places more storage volumes on $SSD1$ servers. Unawareness of the device characteristics leads to an imbalanced state in the left graph. With the performance predictions from accurate consolidation models, the load balancer successfully minimizes the performance difference across the storage hosts, shown in the right graph in Figure \ref{fig:avgl_50}. The statistic of aggregated result indicates that the variance of system-wide average latency is reduced by 82\%. More importantly, the maximum of measured average latency decreases by 52\%, from 12.39ms to 5.95ms, after Serifos load balancing.   

In a multi-tenant environment, cloud clients are interested both for the higher percentiles (i.e. 99) of a performance metric and the average. Therefore, besides the average latency, the 99th percentile of read/write latency are also captured and presented in Figure \ref{fig:read99} and \ref{fig:write99}. Not surprisingly, the I/O load balancer reduces the variance of two measurements by 71\% and 84\% respectively.  
The median and the lowest 50\% of all storage hosts also become much closer in both figures which means there is no overloaded or underloaded host. The majority of clients are able to have more stable performance after Serifos load balancing.

Finally, we present the supported performance SLO setting by two systems in Figure \ref{fig:slo}. When the cloud provider offers performance SLOs, they have to ensure that every workload running on the infrastructure should receive no worse than the SLO in normally 99\% of service time. Otherwise, a penalty will be applied due to the SLA violation. Better performance SLOs are an important consideration when evaluating the workload management system for cloud. Figure \ref{fig:slo} compares the supported SLO value of Cinder and our system for different performance SLO metrics on the same testbed. All three kinds of metric are reduced significantly. After enabling the load balancer, the storage infrastructure is able to support much lower latency SLOs in 99\% of service time. More specifically, 43\% lower on average latency, 32\% lower on maximum read and 63\% lower on maximum write latency.